# Investigation of the phase occurrence and H sorption properties in the $Y_{33.33}Ni_{66.67-x}Al_x$ ($0 \leq x \leq 33.33$) system

Hao Shen[1,2,3,4], Valérie Paul-Boncour[4], Michel Latroche[4], Fermin Cuevas[4], Ping Li[3], Huiping Yuan[1,2], Zhinian Li[1,2], Junxian Zhang[4*] and Lijun Jiang[1,2*]

[1] GRINM Group Co., Ltd., 100088, China, National Engineering Research Center of Nonferrous Metals Materials and Products for New Energy

[2] GRIMAT Engineering Institute Co., Ltd., 101407, China

[3] Institute for Advanced Materials and Technology, University of Science and Technology, Beijing 100083, China

[4] Univ Paris Est Creteil, CNRS, ICMPE, UMR 7182, 2 rue Henri Dunant, 94320 Thiais, France

**Abstract:** The $Y_{33.33}Ni_{66.67-x}Al_x$ system has been investigated in the region $0 \leq x \leq 33.3$. The alloys were synthesized by induction melting. Phase occurrence and structural properties were studied by X-Ray powder Diffraction (XRD). The Al solubility in each phase has been investigated by XRD and Electron Probe Micro-Analysis (EPMA). The hydrogenation properties were characterized by pressure-composition isotherm measurements and kinetic curves at 473 K. For $x = 0$, the binary $Y_{33.33}Ni_{66.67}$ alloy crystallizes in the cubic superstructure with $F\bar{4}3m$ space group and ordered Y vacancies. For $1.67 \leq x \leq 8.33$, the alloys contain mainly Y(Ni, Al)$_2$ and Y(Ni, Al)$_3$ pseudobinary phases; while for $16.67 \leq x \leq 33.33$ they are mainly formed by ternary line compounds $Y_3Ni_6Al_2$, $Y_2Ni_2Al$ and YNiAl. Contrary to the binary $Y_{33.33}Ni_{66.67}$, Y(Ni, Al)$_2$ pseudo-binary compounds crystalize in $C$15 phase (space group $Fd\bar{3}m$) with disordered Y vacancies. The solubility limit of Al in the $C$15 YNi$_{2-y}$Al$_y$ phase is $y = 0.11$ (*i.e.*, $x = 3.67$). The Y(Ni, Al)$_3$ phase changes from rhombohedral (PuNi$_3$-type, $R\bar{3}m$) to hexagonal (CeNi$_3$-type, $P6_3/mmc$) structure for $x$ increasing from 5.00 to 6.67. Upon hydrogenation, the disproportion of $C$15 Y(Ni, Al)$_2$ and losses of crystallinity of YNi and $Y_2Ni_2Al$ are the main reasons causing capacity decay of $Y_{33.33}Ni_{66.67-x}Al_x$ ($0 \leq x \leq 33.33$) alloys upon cycling.

* Correspond author, e-mail addresses: junxian@icmpe.cnrs.fr (Junxian Zhang), jlj@grinm.com (Lijun Jiang)

# 1 Introduction

Hydride forming LaNi$_5$-based $AB_5$-type alloys ($A$ = Rare Earth, $B$ = Transition Metal) are widely used as negative electrode materials for nickel-metal hydride (Ni-$M$H) batteries [1] and hydrogen storage [2]. Recently, ($A$, Mg)Ni-based $AB_n$ alloys ($3 \leq n \leq 3.8$) with stacking structures have been developed for Ni-$M$H batteries [3] with capacity increasing from 1.5 wt.% for $AB_5$-type alloys to 1.8 wt.% for $AB_n$ ($3 \leq n \leq 3.8$)-type alloys [4–7]. $AB_2$-type alloys with $C$15 Laves phase structure possess more tetrahedral sites than $AB_5$ and $AB_n$ ($3 \leq n \leq 3.8$) ones to accommodate hydrogen atoms, thus higher hydrogen absorption capacity is foreseen [8].

Laves phase belongs to the class of Frank-Kasper phases showing topologically close-packed structures. The closest packing of hard spheres can be obtained in $C$15 $AB_2$ alloys, where the components $A$ and $B$ atoms show an ideal radius ratio $r_A/r_B = 1.225$. For intermetallic compounds $AB_2$ ($A$ = rare earth, $B$ = transition metals), the ratio of $r_A/r_B$ is up to 1.4 because of the large atomic radii of lanthanides. As a result of the deviation of the atomic sizes from the ideal ratio, the $A$ atoms exhibit a size contraction and are under stress and $B$ atoms show a reverse trend. Thus, a larger structural instability for light rare earths than for the heavy ones has been observed [9,10]. Among $A$-elements, Y has a medium atomic radius of 1.8 Å (close to that of Gd) [11] but a much lower molar mass. Thus, Y-based $AB_2$-type compounds with $C$15 Laves phase structure should provide better structural stability and larger gravimetric hydrogen absorption capacity than lanthanide-based compounds.

YAl$_2$ crystallizes in the $C$15 structure with a small atomic radius ratio of $r_Y/r_{Al} = 1.26$ close to the ideal value [12,13]. On the contrary, YNi$_2$ ($r_Y/r_{Ni} = 1.45$) does not crystallize in the $C$15 Laves structure but in a superstructure with $A$:$B$ stoichiometry 0.95:2, doubling of the cubic lattice parameter $a$, and lower symmetric space group $F\bar{4}3m$. The superstructure is attributed to the formation of ordered Y vacancies on the 4$a$ sites, which can also be observed for other $A$Ni$_2$ ($A$ = rare earth) compounds [9]. The vacancies allow the relaxation of the stresses caused by the large $r_A/r_{Ni}$ atomic radius ratio Y$_{0.95}$Ni$_2$ absorb a large amount of hydrogen (~3.8 D/f.u.) but suffer from hydrogen-induced amorphization depending on the absorption conditions [14,15]. YNiAl crystallizes in a hexagonal structure $P\bar{6}2m$, it absorbs hydrogen reversibly but its capacity is limited [16]. The Y-Ni-Al system contains several ternary compounds, but not all the compounds have been fully characterized. It is interesting to study this system, focusing on the H$_2$ sorption properties, as it involves only light, non-rare earth elements that could lead to smart materials for energy storage applications.

To our knowledge, no systematic study of the structural and hydrogenation properties for Y$_{33.33}$(Ni, Al)$_{66.67}$ compounds between Y$_{0.95}$Ni$_2$ and YNiAl has been reported. In the present work, we focus on the as-cast alloys, a detailed investigation of the phase occurrence, structural transformation, and hydrogenation properties for Y$_{33.33}$Ni$_{66.67-x}$Al$_x$ (($x$ = 0, 1.67, 3.33, 5.00, 6.67, 8.33, 16.67, 25.00, 33.33) compounds will be investigated.

# 2 Experimental

Alloys with nominal composition Y$_{33.33}$Ni$_{66.67-x}$Al$_x$ ($x$ = 0, 1.67, 3.33, 5.00, 6.67, 8.33, 16.67, 25.00, 33.33) were prepared by induction melting of high-purity metals (Y: 99.9 wt.%, Ni: 99.995 wt.%, Al: 99.99 wt.% from Trillion Metals Co., Ltd.) in high purity argon atmosphere (5N) under a pressure of 0.04 MPa in a water-cooled copper crucible. The ingots were turned over and re-melted three times to ensure homogeneity. Then, parts of the as-cast ingots were mechanically crushed and ground into powder for hydrogenation measurements (< 150 µm) and for X-Ray diffraction (XRD) analysis (< 37 µm).

The global compositions of as-cast alloys were determined by inductively coupled plasma-optical emission spectrometer (ICP). XRD measurements of as-cast alloys were carried with a PANalytical

X-Pert-MRD diffractometer using Cu Kα radiation in the 2θ range of 10º - 90º. Diffraction patterns were collected at 40 kV, 30 mA for the anode of the X-Ray tube and scan rate of 2º min$^{-1}$. Rietveld analyses of the experimental data have been performed with the program Fullprof [17]. The phase composition and microstructure of the alloys were examined by Electron Probe Micro-Analysis (EPMA) in a JEOL-JXA8230 device.

Sorption kinetics and cycling stability of $H_2$ uptake/release reactions were tested in a homemade Sieverts-type apparatus under 2 MPa of hydrogen pressure at 298 K for hydrogenation and dynamic vacuum during 1h at 673 K for dehydrogenation. Pressure-Composition-Isotherms (PCI) curves were measured at 473 K using the Sieverts' method.

## 3 Results

### 3.1 Structural properties

#### 3.1.1 Crystal structure of $Y_{33.33}Ni_{66.67}$ (x = 0)

The Rietveld refinement of the X-ray data for $Y_{33.33}Ni_{66.67}$ is shown in Fig. 1a and the crystallographic data are summarized in Table 1. $Y_{33.33}Ni_{66.67}$ crystallizes in the previously described $Y_{0.95}Ni_2$ superstructure (space group $F\bar{4}3m$) [14,18]. The Rietveld analyses show that small amount of YNi (6 wt. %) and $Y_2O_3$ (2 wt. %) are also present as secondary phases as confirmed by EPMA analyses (Fig. 1b). As already reported, the superstructure is related to a lowering of symmetry (space group $F\bar{4}3m$) due to formation of ordered yttrium vacancies on the 4a site, allowing to release the stress induced by the large $r_A/r_B$ ratio. Lattice parameter and atomic occupation obtained in this work are in good agreement with literature data [14,18]. The observed vacancy ratio (0.6) on the 4a site is slightly smaller than published (0.7) [14], which yields to the stoichiometry $Y_{0.96}Ni_2$ in agreement with the EPMA analyses (Fig. 1b).

TABLE 1 Composition determined by ICP, EPMA and crystallographic data of $Y_{0.96}Ni_2$ from Rietveld refinement.

| | | | | |
|---|---|---|---|---|
| Composition by ICP | | | $Y_{31.93}Ni_{66.67}$ | |
| Composition by EPMA | | | $Y_{0.97(1)}Ni_2$ | |
| Composition by Rietveld | | | $Y_{0.96(1)}Ni_2$ | |
| Atom parameters | Site | Atom | Coordinates | Site occupation factors |
| | 4a | Y1 | x = 0 | 0.40 (1) |
| | 4b | Y2 | x = 1/2 | 1 |
| | 16e | Y3 | x = 0.1028 (3) | 1 |
| | 16e | Y4 | x = 0.6287 (3) | 1 |
| | 24g | Y5 | x = 0.0075 (5) | 1 |
| | 16e | Ni1 | x = 0.3086 (4) | 1 |
| | 16e | Ni2 | x = 0.8136 (5) | 1 |
| | 48h | Ni3 | x = 0.0652 (4) | 1 |
| | | | z = 0.8073 (6) | |
| | 48h | Ni4 | x = 0.0626 (4) | 1 |
| | | | z = 0.3128 (6) | |
| Space group | | | $F\bar{4}3m$ | |
| Cell parameter a (Å) | | | 14.3561 (3) | |

| Number of formula units $Z$ | 64 |
|---|---|
| Volume per formula $V$ (Å$^3$) | 46.23 (1) |
| Phase abundance | 92% |

### 3.1.2 *Phase composition and structure for $Y_{33.33}Ni_{66.67-x}Al_x$ compounds ($1.67 \leqslant x \leqslant 8.33$)*

The relevant XRD patterns and EPMA analyses of the $Y_{33.33}Ni_{66.67-x}Al_x$ ($1.67 \leq x \leq 8.33$) samples are shown in Fig. 2 (Rietveld refinements are given in Fig. S1), the results of Rietveld and EPMA analysis are given in Table 2. The absence of extra peaks related to the superstructure indicates that all Al-substituted compounds crystallize in the Y(Ni, Al)$_2$ ($AB_2$) type $C$15 structure. Meanwhile, Y(Ni, Al)$_3$ ($AB_3$) and YNi ($AB$) appear as secondary phases for $x = 1.67$ and 3.33 respectively. The amounts of secondary phases increase with Al content, and $AB_2$ type structure disappears for $x \geq 8.33$. Indeed, for samples with $x = 6.67$ and 8.33 Y(Ni, Al)$_3$ becomes the main phase (Fig. S1 d and e in supplementary materials). The crystal structure of Y(Ni, Al)$_3$ adopts the *R*-type stacking ($R\bar{3}m$) polymorph for high Al content ($x = 6.67, 8.33$) whereas the *H*-type stacking ($P6_3/mmc$) is favored at low Al content ($1.67 \leq x \leq 5.00$). As observed in the EPMA backscattered electron images (Fig. 2b), the extent of dark gray regions, forming stripes and corresponding to the $AB_3$-type phase rises significantly with Al content, whereas that of light gray regions, attributed to $AB_2$-type phase, decreases. The $AB$-type YNi phase (white areas) is located within the light gray regions and its amount grows gradually with Al content for $3.33 \leq x \leq 8.33$.

Upon substitution, one can observe (Table 2) an increase of the lattice parameters and cell volumes with Al content for both $AB_2$ and $AB_3$ phases. This is consistent with the increasing Al content substituted to Ni inside both $AB_2$ and $AB_3$ phases as measured by EPMA. Furthermore, the yttrium content analyzed by EPMA in all $AB_2$ phases remains sub-stoichiometric (around 0.95), confirming the existence Y-site vacancies, but disordered contrary to the Al-free superstructure phase. On the other hand, the lattice parameters of the $AB$ phase remain almost constant, showing that the Al content in the $AB$ phase is negligible in agreement with EPMA analyses.

TABLE 2 Phase composition determined by ICP, EPMA, phase abundance and crystallographic data from Rietveld refinement for $Y_{33.33}Ni_{66.67-x}Al_x$ ($1.67 \leq x \leq 8.33$) samples.

| Alloys | | $x$ | 1.67 | 3.33 | 5.00 | 6.67 | 8.33 |
|---|---|---|---|---|---|---|---|
| | | ICP | $Y_{31.61}Ni_{64.95}Al_{1.72}$ | $Y_{32.09}Ni_{63.28}Al_{3.39}$ | $Y_{32.23}Ni_{61.71}Al_{4.96}$ | $Y_{32.04}Ni_{60.03}Al_{6.64}$ | $Y_{32.98}Ni_{58.51}Al_{8.16}$ |
| Phases | Y (Ni, Al)$_2$ ($C$15, $Fd\bar{3}m$) | EPMA (±0.02) | $Y_{0.94}Ni_{1.94}Al_{0.05}$ | $Y_{0.95}Ni_{1.91}Al_{0.09}$ | $Y_{0.96}Ni_{1.89}Al_{0.15}$ | $Y_{0.95}Ni_{1.88}Al_{0.12}$ | - |
| | | $a$ (Å) | 7.1822 (2) | 7.1985 (2) | 7.2073 (2) | 7.2072 (4) | - |
| | | $V$ (Å$^3$) | 370.49 | 373.00 | 374.39 | 374.44 | - |
| | | Abundance (wt.%) | 89% | 76% | 54% | 24% | |
| | Y (Ni, Al)$_3$ ($R\bar{3}m \leqslant 5.00$ $P6_3/mmc \geqslant 6.67$) | EPMA (±0.02) | - | $Y_{1.03}Ni_{2.85}Al_{0.15}$ | $Y_{1.03}Ni_{2.79}Al_{0.21}$ | $Y_{1.03}Ni_{2.70}Al_{0.30}$ | $Y_{1.03}Ni_{2.61}Al_{0.39}$ |
| | | $a$ (Å) | 4.988 (1) | 5.0019 (9) | 5.0208 (3) | 5.0457 (3) | 5.0667 (3) |
| | | $c$ (Å) | 24.427 (1) | 24.4410 (8) | 24.4006 (4) | 16.2664 (1) | 16.2349 (1) |
| | | $V$ (Å$^3$)/$AB_3$ | 175.58 | 176.50 | 177.57 | 179.34 | 180.48 |
| | | Abundance (wt.%) | 5% | 10% | 19% | 46% | 63% |
| | YNi ($Pnma$) | EPMA (±0.02) | - | YNi$_{1.00}$ | YNi$_{1.02}$Al$_{0.01}$ | YNi$_{1.04}$Al$_{0.02}$ | YNi$_{0.99}$ |
| | | $a$ (Å) | | 7.121 (1) | 7.146 (1) | 7.137 (2) | 7.1365 (8) |
| | | $b$ (Å) | | 4.1377 (1) | 4.1394 (1) | 4.1296 (1) | 4.1280 (1) |

| | | | | | | |
|---|---|---|---|---|---|---|
| | | c (Å) | 5.5134 (1) | 5.4938 (2) | 5.5096 (1) | 5.5140 (1) |
| | | V (Å³) | 162.45 | 162.51 | 162.38 | 162.44 |
| | | Abundance (wt.%) | 5% | 12% | 16% | 23% |
| | EPMA (±0.02) | - | $Y_2Ni_{2.48}Al_{0.18}$ | $Y_2Ni_{2.42}Al_{0.40}$ | $Y_2Ni_{2.42}Al_{0.42}$ | $Y_2Ni_{2.32}Al_{0.54}$ |
| | a (Å) | | 4.1615 (1) | 4.153 (1) | 4.154 (1) | 4.1545 (7) |
| $Y_2Ni_2Al$ | b (Å) | | 5.3478 (6) | 5.3374 (1) | 5.3323 (1) | 5.3385 (1) |
| (Immm) | c (Å) | | 8.2247 (8) | 8.3008 (1) | 8.3015 (1) | 8.3049 (1) |
| | V (Å³) | | 183.05 | 184.02 | 183.88 | 184.20 |
| | Abundance (wt.%) | | 4% | 13% | 12% | 12% |
| $Y_2O_3$ ($Ia\bar{3}$) | Abundance (wt.%) | 6% | 5% | 2% | 2% | 2% |

### 3.1.3 Phase composition and structure for $Y_{33.33}Ni_{66.67-x}Al_x$ ($16.67 \leq x \leq 33.33$) compounds

Fig. 3a shows the XRD patterns for $16.67 \leq x \leq 33.33$. For $x = 16.67$, the XRD pattern can be indexed with cubic $Y_3Ni_6Al_2$ ($Ca_3Ag_8$-type, $Im\bar{3}m$) [19] and orthorhombic $Y_2Ni_2Al$ ($W_2CoB_2$-type, $Immm$) [20] structures beside a small quantity (< 5 wt.%) of $Y_2O_3$. For the sample with $x = 25.00$, the hexagonal YNiAl phase ($Fe_2P$-type, $P\bar{6}2m$) [16] appears as the main phase with some $Y_2Ni_2Al$ and Y(Ni, Al)$_5$ ($P6/mmm$) secondary ones. For the sample with $x = 33.33$, YNiAl is almost single-phase with very small amount (≤ 3 wt. %) of $Y_2Ni_2Al$ and Y(Ni, Al)$_5$ (Fig. S1 f, g, h in the supplementary materials). The lattice parameters, cell volumes and phase amount obtained from Rietveld analyses of the XRD patterns are given in Table 3. For this Al content region, the $AB$, $AB_2$, and $AB_3$ phases are not detected. EPMA back scattered electron images and corresponding analysis of the as-cast alloys with $x = 16.67, 25.00, 33.33$ are shown in Fig. 3b and summarized in Table 3. The measured compositions for $Y_2Ni_2Al$ and YNiAl phases agree with previously reported ones, though there are deviations on Ni and Al contents from the stoichiometric composition of $Y_2Ni_2Al$. These deviations are accompanied by variations of the lattice parameters (table 3), suggesting a homogeneity domain for this phase rather than a line compound.

TABLE 3 Phase composition determined by ICP, EPMA and crystallographic data from Rietveld refinement for $Y_{33.33}Ni_{66.67-x}Al_x$ ($16.67 \leq x \leq 33.33$) samples.

| Alloys | | x | 16.67 | 25.00 | 33.33 |
|---|---|---|---|---|---|
| | | ICP | $Y_{32.41}Ni_{50.33}Al_{16.34}$ | $Y_{32.13}Ni_{41.70}Al_{24.97}$ | $Y_{32.45}Ni_{33.72}Al_{32.95}$ |
| | | EPMA (± 0.02) | $Y_3Ni_{5.91}Al_{1.92}$ | $Y_3Ni_{5.34}Al_{1.95}$ | - |
| | $Y_3Ni_6Al_2$ | a (Å) | 8.9329 (1) | 8.9360 (2) | |
| | ($Im\bar{3}m$) | V (Å³) | 712.82 | 713.57 | |
| Phases | | Abundance (wt.%) | 56% | 26% | - |
| | | EPMA (± 0.02) | $Y_2Ni_{2.28}Al_{0.70}$ | $Y_2Ni_{2.06}Al_{0.94}$ | - |
| | | a (Å) | 4.1593 (2) | 4.1680 (2) | 4.165 (2) |
| | $Y_2Ni_2Al$ | b (Å) | 5.3977 (1) | 5.4220 (3) | 5.509 (3) |
| | (Immm) | c (Å) | 8.3450 (2) | 8.3706 (5) | 8.3413 (6) |
| | | V (Å³) | 187.35 | 189.17 | 191.44 |

| | | | | |
|---|---|---|---|---|
| | Abundance (wt.%) | 42% | 26% | 1% |
| YNiAl ($P\bar{6}2m$) | EPMA (± 0.02) | - | YNi$_{1.01}$Al$_{1.00}$ | YNi$_{0.95}$Al$_{0.98}$ |
| | $a$ (Å) | | 7.0306 (1) | 7.0410 (2) |
| | $c$ (Å) | | 3.8327 (1) | 3.8411 (3) |
| | $V$ (Å$^3$) | | 164.06 | 164.92 |
| | Abundance (wt.%) | - | 42% | 95% |
| Y(Ni, Al)$_5$ ($P6/mmm$) | | - | - | - |
| | $a$ (Å) | | 5.0298 (3) | 5.058 (1) |
| | $c$ (Å) | | 4.0711 (4) | 4.0827 (2) |
| | $V$ (Å$^3$) | | 89.20 | 90.48 |
| | Abundance (wt.%) | - | 5% | 2% |
| Y$_2$O$_3$ ($Ia\bar{3}$) | Abundance (wt.%) | 2% | 1% | 2% |

3.2  Hydrogenation properties

The hydrogenation kinetics of all alloys ($0 \leq x \leq 33.33$) for the first and the fourth hydrogen absorption cycles at 298 K under 2 MPa H$_2$ are shown in Fig. 4. For the first cycle (Fig. 4a), after nearly 20 hours, the hydrogen storage capacity of Y$_{33.33}$Ni$_{66.67}$ attains 1.74 wt. %, which corresponds to the formation of the metal hydride YNi$_2$H$_{3.6}$. The hydrogen absorption kinetics are much faster at the fourth cycle with full uptake in 5 min (Fig. 4b). The hydrogen absorption capacity decays significantly after the first cycle, then stabilizes at 1.08 wt. % for Y$_{33.33}$Ni$_{66.67}$. For Al-substituted compounds ($x > 0$), the kinetics of the first absorption is significantly improved as compared to the Al-free sample. The maximum capacity is reached in less than one hour for $x > 3.33$. As shown in Fig. 4c and d, the maximum capacity is as high as 1.8 wt. % for $x = 1.67$ and then decreases monotonously with Al content to reach 0.75 wt.% for YNiAl ($x = 33.33$). After four cycles, the samples with $x = 1.67, 3.33, 5.00, 8.33$ show similar stable capacity around 1.2 wt.%, higher than for the binary Y$_{33.33}$Ni$_{66.67}$ (1.08 wt.%). For higher aluminum substitution, the capacity is lowered to 0.75 wt.% (Fig. 4d). It is noticed that YNiAl shows more stable hydrogen absorption behavior although its capacity is low (0.63 wt.%).

Fig. 5 shows the *P-C* Isotherms (PCI) of the alloys ($x = 0, 1.67, 3.33, 5.00, 8.33, 33.33$) during the first cycle at 473K. Two plateaus can be identified for Al-free Y$_{33.33}$Ni$_{66.67}$ with equilibrium pressures at 0.01 and 1MPa, respectively.

For alloy with $x = 1.67$, the PCI shows similar profile to that of Y$_{33.33}$Ni$_{66.67}$ ($x = 0$) with higher pressure for the first plateau (0.03 MPa) and sloping behavior for the second one. For $x=0$, the two-plateau behavior shows clearly, while for $x = 3.33$ and 5, the first plateau remains present while second one is a sloped branch. At low hydrogen content, the PCI show sloping branches and narrow plateaus at 0.03 MPa between 0.4 H/M and 0.7 H/M, then a sloping curve raising up to 1.2 H/M is observed (Fig. 5a). For $x = 8.33$ (Fig. 5b), up to 0.4 H/M the equilibrium pressure is too low, *i.e.* below 10$^{-5}$ MPa, to be determined by our Sieverts rig. However, a sloping plateau ranging from 0.5 to 1.2 H/M and with pressure of ~ 1 MPa appears, showing good reversibility and very small hysteresis. For $x = 33.33$, the PCI shows a sloping branch and large hysteresis at 473K.

To determine whether structural changes are induced by hydrogenation, the XRD patterns of hydrogenated $Y_{33.33}Ni_{66.67}$ (298K under 2 MPa of $H_2$) and dehydrogenated samples $0 \leq x \leq 33.33$ (dynamic vacuum at 673K) have been analyzed. For $Y_{33.33}Ni_{66.67}$ (Fig. S2 a, b and Table 4), the Rietveld analyses of the hydrogenated sample show a cubic $Y_{0.96}Ni_2H_z$ structure with a lattice expansion $\Delta V/V = 18.5$ % in agreement with the literature [14]. After dehydrogenation under dynamic vacuum at 673K, the XRD pattern can be indexed with $YNi_3$ and $YH_2$, indicating a disproportionation reaction of the hydride upon hydrogenation cycling. Similar behavior is observed for $1.67 \leq x \leq 8.33$. The pristine alloys contain $C15$ $Y(Ni,Al)_2$, $(Y(Ni,Al)_3$ and few $YNi$, whereas after hydrogenation/dehydrogenation process, the $C15$ phase is no more observed, but the patterns are indexed with $Y(Ni, Al)_3$ and $YH_2$ (Fig. 6a). $Y(Ni,Al)_3$ crystallizes in rhombohedral structure up to $x = 5.0$ and in hexagonal structure for $x \geq 6.67$. These results indicate a disproportion of $Y(Ni,Al)_2$ $C15$ phase into $AB_3$ phase and Y hydride upon hydrogenation/dehydrogenation. For higher Al content ($x = 16.67, 25.00$), $Y_3Ni_6Al_2$ and $YNiAl$ phases are still observed after one hydrogenation cycle, but $Y_2Ni_2Al$ disappears and minor $YH_2$ is formed (Fig. 6b). For $x = 33.33$ ($YNiAl$), neither structural changes, nor peak broadening are observed in the XRD patterns upon cycling, in agreement with the results reported by Kolomiets [21]. For the alloys containing initially $Y_2Ni_2Al$ and $YNi$, these two phases cannot be detected after one hydrogenation/dehydrogenation cycle.

TABLE 4 Crystallographic data from Rietveld refinement for $Y_{33.33}Ni_{66.67}$ sample upon hydrogenation and dehydrogenation.

| Condition | Phase | X-ray refined parameters | | | | |
|---|---|---|---|---|---|---|
| | | $a$ (Å) | $c$ (Å) | $V$ (Å$^3$) | $\triangle V/V$ (%) | Occupation (Y: 4$a$ site) |
| Hydrogenated | $Y_{0.95}Ni_2D_{2.6}$ ($F\bar{4}3m$) [14] | 15.113 (1) | | 3451.7 (4) | 16.7% | 0 |
| | $Y_{0.96}Ni_2H_z$ ($F\bar{4}3m$) | 15.192 (1) | | 3506.4 (5) | 18.5% | 1.75 (1) |
| Dehydrogenated | $YNi_3$ ($R\bar{3}m$) | 4.9696 (8) | 24.3253 (4) | 520.2 (1) | | |
| | $YH_2$ ($Fm\bar{3}m$) | 5.210 (1) | | 141.4 (1) | | |

## 4 Discussion

As the system $Y_{33.33}Ni_{66.67-x}Al_x$ shows a large number of phases and structural transformations in the domain $0 \leq x \leq 33.33$, the formation, structural evolution and Al solubility properties of $Y_{0.95}Ni_2$, $Y(Ni, Al)_2$, $Y(Ni, Al)_3$, $Y_3Ni_6Al_2$, $YNiAl$, $Y_2Ni_2Al$ will be considered first, then the hydrogenation properties of the alloys will be discussed.

### 4.1 Structural properties

#### a. $Y_{0.95}Ni_2$ phase with superstructure

$Y_{33.33}Ni_{66.67}$ crystallizes in the superstructure described in the lower symmetry $F\bar{4}3m$ space group with Y:Ni stoichiometry 0.95:2. This superstructure is characterized by ordered Y vacancies on the 4$a$ site and a doubling of the lattice parameter of the Laves phase ($a = 14.35$ Å). For $ANi_2$ ($A$ = Y, rare earth) binary compounds, the instability of the $C15$ Laves phase structure ($Fd\bar{3}m$) is usually attributed to the large atomic radius ratio $r_A/r_{Ni}$ (up to 1.45) above the geometrical ideal one (1.225). The introduction of $A$ vacancies allows relaxing constraints thus stabilizing the structure [9]. Therefore, the formation of $Y_{0.95}Ni_2$ with superstructure occurs instead of the $C15$ Laves phase for $x = 0$. To allow chemical balance, some $YNi$ phase is formed. We observed 6 wt.% YNi and 2 wt.% yttrium oxide. This value closely agrees to the calculated 6.8 wt.% YNi expected from the stoichiometric reaction: $YNi_2 \Rightarrow 0.952\ Y_{0.95}Ni_2 + 0.095 YNi$.

*b. $Y_{1-v}(Ni, Al)_2$ - C15 phase*

Fig. 7 shows the phase abundance as the function of the nominal Al content for $0 \leq x \leq 8.33$, where the $AB_2$ amount decreases monotonously down to vanish at $x = 8.33$. For the sake of comparison, $Y_{0.95}Ni_2$ is accounted as an $AB_2$ phase even if it is off-stoichiometric with superlattice structure. As the nominal aluminum content increases to $x = 1.67$, a reversed structural transformation for Al-free $YNi_2$ ($x = 0$) from superstructure to $C15$ Laves phase structure for $Y(Ni, Al)_2$ solid solution ($0 < x \leq 5.00$) can be observed. Al-containing alloys are multiphase but the main phase is indeed the $AB_2$ one for $x \leq 5.00$. Nevertheless, EPMA shows a systematic sub-stoichiometry for yttrium indicating the formation of disordered Y vacancies. Therefore, the substitution of Ni by Al favors the formation of the $C15$ structure without Y-vacancy ordering compared to the binary $Y_{0.95}Ni_2$. Similar behavior was observed with $B$ = Cu or Fe in the $Y_{0.95}Ni_{2-x}B_x$ system, showing for specific substitution ratio the $C15$ structure with disordered Y vacancies, *i.e.* without superstructure [22].

Above $x = 5.00$, the $AB_2$ phase becomes secondary and disappears at $x = 8.33$ to the profit of $AB_3$ that increases up to 63wt.%.

Fig. 8 shows the evolution of the cubic lattice constant and cell volume of the $AB_2$-type $YNi_{2-y}Al_y$ phase with nominal Al content (bottom-axis $x$) and Al content measured EPMA (top-axis $y$). The lattice constant increase linearly with Al as expected for a solid solution following the Vegard's law [23], except at low $x$ values (0 to 1.67) for which the structural change occurs at nearly constant volume. For nominal contents $5.00 \leq x \leq 6.67$, both the Al content ($y = 0.11$ and $0.12$, from EPMA, respectively) and the lattice constant ($a = 7.2073$ Å) of the $AB_2$-type phase remain constant. This indicates that the solubility limit of Al in $YNi_{2-y}Al_y$ $C15$ structure is reached and is limited to $y \approx 0.12$.

*c. $Y(Ni, Al)_3$ phase with stacking structure*

As shown in Fig. 7, the $AB_3$ emerges as a second phase when Al content reaches $x = 1.67$ and becomes the main phase, increasing to 63 wt. %, for $x = 8.33$. It is interesting to note the structural change for $Y(Ni, Al)_3$ from rhombohedral (PuNi$_3$-type, $3R$) for $x \leq 5.00$ (*i.e.* $YNi_{2.8}Al_{0.2}$) to hexagonal (CeNi$_3$-type, $2H$) for $x \geq 6.67$ (*i.e.* $YNi_{2.7}Al_{0.3}$; see Table 1 and Fig.9). Similar results have been reported for La(Ni, Mn)$_3$ phase [24]. Indeed, Denys *et al.* found that substituting Ni for Mn in the LaNi$_{3-w}$Mn$_w$ system induces a structural transition from rhombohedral ($3R$) to hexagonal ($2H$) for $w > 0.1$. It can be hypothesized that Ni substitution by larger atoms like Al or Mn promotes the formation of the hexagonal structure rather than the rhombohedral one. Similar results have been also reported for the $A_2Ni_7$ ($A$ = Rare Earth) systems with stacking structure and they have been related to geometrical effects [25–28].

Fig. 9 shows the lattice constant and cell volume evolutions for $Y(Ni, Al)_3$ with nominal Al content $x$ (bottom axis, solid line) and Al content $y$ obtained by EPMA (top axis, dashed line). For the sake of comparison between $R$ and $H$ crystal structures, the lattice constant $c$ of the hexagonal structure is multiplied by 3/2. One can notice that the lattice constant $a$ increases almost linearly with $x_{Al}$. The lattice constant $c$ tends to decrease, which means that the lattice expands along the *basal* plane and contracts slightly along the $c$ axis. This phenomenon can be explained presuming that Al behaves like Mn in LaNi$_{3-w}$Mn$_w$ where Mn substitutes to Ni only in the $AB_5$ slab [24]. This leads to an expansion of this slab, a reduced mismatch between the $AB_5$ and $A_2B_4$ slabs, and finally causes a lattice expansion in the $ab$ plan and shrinkage of the $c$ direction.

For $x$ = 6.67 and 8.33, the EPMA compositions for $AB_3$ are $Y_{1.03}Ni_{2.70}Al_{0.30}$ and $Y_{1.03}Ni_{2.61}Al_{0.39}$ respectively, very close to the reported $Y_3Ni_8Al$ ternary phase [29]. This latter phase crystallizes in a $CeNi_3$-type stacking structure with ordered Al occupation on the 2$a$ site. It can be represented as a stacking of $Y_2Ni_3Al$ ($AB_2$ type structure) and $YNi_5$ ($AB_5$-type structure) along the $c$-axis. Indeed, the stacking structure stability is related to the (mis)match of both $AB_2$ and $AB_5$ slabs. As the $AB_2$ slab is originally larger than the $AB_5$ one, only partial substitutions, either of $A$ by smaller atoms (like Mg) in the $AB_2$ [24,30] or/and of $B$ by larger ones (Mn or Al) in the $AB_5$ are reported to stabilize the stacking structure and improve their hydrogenation, forming a flatter plateau and higher reversibility [31–33]. From a first-principle study, the $Y_3Ni_8Al$ structure was determined and the calculation shows that the Al atom occupies the Wyckoff site 2$b$ inside the $AB_5$ slab [34]. Another experimental study on the Y-Ni-Ga system showed that Ga atoms, similar to Al, occupy the 2$c$ site in $AB_5$ slab [35]. Our Rietveld analyses also show that Al occupies preferably the sites in $AB_5$ slab (Wyckoff sites 2$b$ and 2$c$) and at the border between $AB_2$ and $AB_5$ slab (Wyckoff site 12$k$). So, from our results, $Y_{1.03}Ni_{2.70}Al_{0.30}$ and $Y_{1.03}Ni_{2.61}Al_{0.39}$ phases are solid solutions with Al occupying preferably the $AB_5$ slab $B$ sites in agreement with literature [24,30]. The result of Rietveld analyses is show in Figure S3 by comparison with Al atom occupying only the 2$a$ site.

*d. YNi and $Y_2Ni_2Al$ phases*

As can be seen from Fig. 7, for $x \leq 8.33$, beside the two main phases $AB_2$ and $AB_3$, two minor phases YNi and $Y_2Ni_2Al$ are formed for the compensation of additional Y induced by the formation of the $AB_3$-type phase. Interestingly, YNi is very poor in Al (see Al composition detected by EPMA from table 2) and it can be considered as a binary compound with very low Al solubility. $Y_2Ni_2Al$ is a ternary compound reported in the literature [36] but not shown in the ternary phase diagram published by Ferro *et al.*[37]. In the present work, $Y_2Ni_2Al$ shows variable Al content. On increasing the nominal Al content in $Y_{33.33}Ni_{66.67-x}Al_x$ samples, the Al content of the $Y_2Ni_2Al$ phase increases from 3.6 at. % to 20 at. % Al along a line joining $Y_2Ni_{2.5}Al_{0.2}$ for $x$ = 1.67 to $Y_2Ni_2Al$ for $x$ = 25.00.

*e. YNiAl phase*

As can be seen Fig. 10, for $x$ values larger than 16.67, the ternary line compounds $Y_3Ni_6Al_2$ and YNiAl are formed instead of $AB_2$ or $AB_3$ pseudo-binary phases. Moreover, the abundances of $Y_3Ni_6Al_2$ and $Y_2Ni_2Al$ decrease rapidly as the Al content approaches the ternary composition YNiAl. In agreement with the Y-Ni-Al phase diagram [38], the composition of the observed compounds are located in the region where the three ternary phases coexist. For $x$ = 33.33, expectedly the alloy is almost single phase YNiAl with small amount of oxide. As a consequence, in the studied region, there are two line compounds, YNi and YNiAl. For $YNi_{2-x}Al_x$ $AB_2$ and $YNi_{3-x}Al_x$ $AB_3$ phases, the solution limit of Al is $x$ = 0.11 and $x$ = 0.4, respectively. For $Y_2Ni_2Al$, the phase can form with Al content as low as $x$ = 0.2.

*4.2 Hydrogenation properties*

*a. The Pressure Composition Isotherm*

For $x$ = 0, the hydrogenation and dehydrogenation properties can be attributed to $Y_{0.95}Ni_2$ ($F\bar{4}3m$) as it is the major phase (~ 92 wt.%). In the literature, the crystal structure of $Y_{0.95}Ni_2D_{2.6}$ prepared at room temperature with 1 bar of deuterium pressure has been reported [14]. Another work [39] reported a continuous increase of the pressure-composition curve in the range $10^{-3}$ to 1 MPa, without plateau pressure at room temperature. For the first time, we show here the complete PCI of $Y_{0.95}Ni_2$ at 473 K. This is in good agreement with the results of reference [14], which state that high a large gap of pressure is necessary to increase the hydrogen content significantly. It probably means the end of the first plateau and the beginning of the second one. We consider that the two-plateau of the PCI correspond to the two hydrides $Y_{0.95}Ni_2H_{2.6}$ and $Y_{0.95}Ni_2H_3$. Then, $Y_{0.95}Ni_2D_{2.6}$ can be identified as a first hydride whereas the crystal structure of the second hydride at the end of the second plateau remains unknown but could be amorphous as observed for other $R$Ni$_2$ hydrides . For confirmation, in-situ neutron diffraction studies under hydrogen (deuterium) pressure will be valuable.

For $x = 1.67$, $Y_{0.95}(Ni, Al)_2$ ($Fd\bar{3}m$-C15) is the major phase (~ 90 wt.%), which allows a direct comparison of the *PCI* with $x = 0$. At 473 K, we observe that for $x = 1.67$, the pressure of the low-pressure plateau (Fig. 5a) is higher than that of $x = 0$ though the lattice parameter of the former, $a = 7.182$ Å, is larger than the latter, $a/2 = 7.178$ Å. This contradicts the geometrical rule obeyed by many metal-hydrogen systems [40–42] showing that the lower the cell parameter, the higher the pressure of the plateau. Such contradiction is commonly attributed to the predominancy of electronic versus geometric effects [43,44]. For $x = 3.33$ and 5.00, the alloys contain other phases and the PCI is expected to reflect the contribution of each component (Figure 5a). The sloped branch at low pressure could be assigned to YNi and $Y_2Ni_2Al$ phases. For $x = 8.33$, the second plateau at 1 MPa and 0.7 H/M can be attributed to $YNi_{2.6}Al_{0.4}$ ($AB_3$ phase) (according to EPMA results, see Fig. 5b), which shows very good reversibility. Accordingly, it will be worth to synthesis the single-phase compound $YNi_{2.6}Al_{0.4}$ for further investigation as hydrogen storage material.

At higher Al content ($x \geqslant 16.67$), the PCI show very sloping behavior and lower absorption capacity due to the presence of the phases $Y_2Ni_2Al$, $Y_3Ni_6Al_2$ and YNiAl.

*b. Kinetics*

During absorption, the hydrogenation can be described as a sequential process comprising physisorption and chemisorption of hydrogen molecules at the surface, dissociation of dihydrogen, surface penetration of hydrogen atoms into the solid, diffusion of hydrogen atoms in the bulk and nucleation and growth of the hydride phase. Some models exist to identify the rate-determining steps [45]. For $Y_{33.33}Ni_{66.67}$, the sorption curve exhibits a sigmoidal shape (Figure 4a) indicating that the limiting step is the nucleation and growth of hydride nuclei [46]. The hydrogenation kinetics for $1.67 \leq x \leq 8.33$ are faster compared to Al-free alloy. As for the Al-content ranging from $x = 1.67$ to 5.00, both rates of hydride nucleation and growth increases. This can be attributed to the presence of the secondary phases Y(Ni, Al)$_3$, YNi and $Y_2Ni_2Al$ which may play a catalytic role on the dihydrogen dissociation process. The kinetic of alloys with higher Al ($x = 25.00, 33.33$) contents is slower than that of $x = 16.67$, all Al-contain alloys show better kinetic than the Al-free alloy.

*c. Reversible capacity*

Regarding the maximum hydrogen sorption capacity (Fig. 11), it increases slightly for $x = 1.67$ as compared to the Al-free alloy, which is attributed to the decrease of the molar weight by Al substitution in the $AB_2$ phase, thus augmenting the mass capacity. When further increasing Al content, the maximum absorption capacity decreases due to the formation of secondary phases $Y_2Ni_2Al$, $Y_3Ni_6Al$ and YNiAl, which do not absorb large quantity of hydrogen.

All alloys show a decay of their hydrogen absorption capacity between the first and the second cycle (Fig. 4c), but the reason is different for each alloy.

For $x = 0$, XRD patterns of the dehydrogenated samples can be indexed with $YNi_3$ and $YH_2$ (Fig. S2 and Fig. 6). The capacity decay is mainly attributed to the disproportionation of $Y_{0.95}Ni_2H_z$. As a matter of fact, upon hydrogenation $AB_2$ ($A$ = Y, rare earth, $B$ = transition metals) binary compounds with $C15$ Laves phase structure suffer from hydrogen-induced amorphization (HIA) or hydrogen-induced disproportionation [10,47]. Indeed, $Y_{0.95}Ni_2$ keeps the structural characteristics of the $C15$ Laves structure but needs ordered Y vacancies to stabilize. Despite this effect, any volume expansion induced by hydrogenation leads to interatomic distance distribution and weakening of atomic bounds [9], thus easier occurrence for HIA.

For $1.67 \leq x \leq 8.33$, the XRD patterns of the dehydrogenated samples can be indexed with $Y(Ni, Al)_3$ and $YH_2$ phases. This result indicates that the same hydrogen-induced disproportionation occurred for all present phases except $AB_3$-type $Y(Ni, Al)_3$ phase. Neither YNi nor $Y_2Ni_2Al$ phase contribute to the reversible hydrogen absorption capacity as for samples with Al content ranging between $3.33 \leq x \leq 8.33$, upon hydrogen absorption and desorption the diffraction peaks of YNi and $Y_2Ni_2Al$ phases cannot be detected. However, YNi could absorb approximately 3 H/f.u. under 1.1MPa of hydrogen pressure at room temperature to form a hydride $YNiH_3$ [48]. $YNiH_3$ is instable as upon hydriding it transforms into yttrium hydride and Ni. For $Y_2Ni_2Al$ phase, though there are no studies on its hydrogenation properties, some studies for the isostructural intermetallic compounds $A_2Ni_2Al$ ($A$ = Gd, Er, Lu) showed that hydrogenation results in strong anisotropic cell expansion [20,49], which can eventually yield HIA.

The reversible capacity shows significant increase for Al contain alloys up to $x = 8.33$, which is ensured by the $Y(Ni, Al)_3$ phases after the first cycle. This indicates that Al-containing $Y(Ni, Al)_3$ provides higher hydrogen capacity than the binary $YNi_3$ one. For alloys with $x = 16.67$, 25.00 and 33.33, the XRD patterns after dehydrogenation show diffraction peaks belonging to $Y_3Ni_6Al_2$ and YNiAl (Fig. 6b). This means that these two phases recover their crystal structure after dehydrogenation, which agrees with previous experimental results about the hydrides $A_3Ni_6AlGaH_z$ ($A$ = Y, Gd, Dy) [19] and $ANiAlH_z$ ($A$ = Y, Gd, Tb, Dy, Ho and Er) [21]. Fig 12 show the $Y_2Ni_2Al$ phase abundance and capacity decay evolutions as a function of $x_{Al}$ in the region $16.67 \leq x \leq 33.33$. They display similar tendency, indicating that HIA of $Y_2Ni_2Al$ is the main reason for the capacity decay between $16.67 \leq x \leq 33.33$ Al content. The low reversible capacity is caused by the low hydrogen capacity of YNiAl and $Y_3Ni_6Al_2$.

## 5  Conclusions

We investigated systematically the phase occurrence for $Y_{33.33}Ni_{66.67-x}Al_x$ in the range of $0 \leq x \leq 33.33$. We established that the Al substitution limit inside the $C15$ $YNi_{2-y}Al_y$ phase is $y = 0.11$. When increasing $x_{Al}$ from 5.00 to 6.67, the $AB_3$-type $Y(Ni,Al)_3$ phase occurs. Its structure switches from rhombohedral (PuNi$_3$-type, $3R$) to hexagonal (CeNi$_3$-type, $2H$) while the Al content increases, due to geometric effects. The Al solution limit for $YNi_{3-y}Al_y$ is estimated around $y = 0.4$. The reported ordered $Y_3Ni_8Al$ ternary phase [29] is not observed but $YNi_{2.6}Al_{0.4}$ (with close composition, i.e. $Y_3Ni_{7.8}Al_{1.2}$) crystallizes as a pseudo-binary in the same hexagonal structure, without any ordering of Al atoms in the $2d$ sites.

Al substitution enhances the hydrogen absorption kinetics of $Y_{33.33}Ni_{66.67-x}Al_x$ and improves the reversible hydrogen sorption capacity. The $Y_{0.95}Ni_2$ (superstructure phase), $Y_{0.95}(Ni, Al)_2$ ($C15$ structure), YNi and $Y_2Ni_2Al$ are not stable upon hydrogen absorption/desorption. For alloys with Al content lower than 8.33, the reversible capacity is ensured by the $AB_3$ phase which formed after the first cycle. For higher Al content, the reversible capacity comes from ternary phases $Y_3Ni_6Al_2$ and YNiAl. The $AB_3$-type $Y_{1.03}Ni_{2.61}Al_{0.39}$ phase shows interesting hydrogenation properties. A single-phase compound will be worth to synthesize to further investigate its hydrogen sorption properties.

**Acknowledgement**

This work was supported by the National Key R&D Program of China [2019YFE0103600] and PHC CAI YUANPEI project (44027WH).

**Figure caption:**

Fig. 1. Rietveld refinement of X-ray diffraction pattern (a) and EPMA back scattered electron image (b) of $Y_{33.33}Ni_{66.67}$ ($x = 0$).

Fig. 2. X-ray diffraction patterns (a) and EPMA back scattered electron images (b) for $0 \leq x \leq 8.33$: a - $AB_2$ phase (light gray); b - $AB_3$ phase (dark gray); c, d - $AB$ and $Y_2Ni_2Al$ phases (white). The deep-black regions are attributed to Y oxide and some cracks produced during grinding.

Fig. 3. X-ray diffraction patterns (a) and EPMA back scattered electron images (b) for $Y_{33.33}Ni_{66.67-x}Al_x$ ($16.67 \leqslant x \leqslant 33.33$) samples.

Fig. 4. Time dependence of the hydrogen absorption for $0 \leqslant x \leqslant 33.33$ alloys at the first (a) and fourth (b) cycle (hydrogenation conditions: $P_{H2} = 2MPa$, 298K, dehydrogenation conditions: dynamic vacuum, 673K); The hydrogen absorption capacity as function of cycle number for $0 \leqslant x \leqslant 33.33$ (c) under same conditions.

Fig. 5. *P-C* isotherms measured at 473K for $Y_{33.33}Ni_{66.67-x}Al_x$ ($0 \leq x \leq 33.33$) samples.

Fig. 6. X-ray diffraction patterns for $Y_{33.33}Ni_{66.67-x}Al_x$ samples ($0 \leq x \leq 8.33$ (a), $16.67 \leq x \leq 33.33$ (b)) after one hydrogenation/dehydrogenation cycle.

Fig. 7. Phase abundance as the function of the nominal Al content for $0 \leq x \leq 8.33$. $AB_3$ phase exhibits PuNi$_3$-type structure below $x = 5.00$ and CeNi$_3$-type structure above this value, the dashed line is a guide for eyes indicating the structure change of $AB_3$ phase.

Fig. 8. Lattice constant *a* and cell volumes *V* of Y(Ni, Al)$_2$ as function of nominal Al content *x* for $Y_{33.33}Ni_{66.67-x}Al_x$ alloys (below) as well as Al content in the Y(Ni, Al)$_2$ phases obtained by EPMA (top).

Fig. 9. Lattice constants *a, c* and cell volumes *V* of Y(Ni,Al)$_3$ phases as function of nominal Al content *x* in $Y_{33.33}Ni_{66.67-x}Al_x$ alloys (bottom) as well as Al content obtained by EPMA in the YNi$_{3-y}$Al$_y$ phase (top).

Fig. 10. Phase abundance of $Y_3Ni_6Al_2$, $Y_2Ni_2Al$, YNiAl and Y(Ni, Al)$_5$ for $16.67 \leq x \leq 33.33$.

Fig. 11 Maximum and reversible capacity as function of Al content in $Y_{33.33}Ni_{66.67-x}Al_x$ ($0 \leqslant x \leqslant 33.33$).

Fig. 12. Evolution of $Y_2Ni_2Al$ abundance and capacity decay between cycle 1 and 4 as a function of the Al content ($16.67 \leqslant x \leqslant 33.33$).